\documentclass[aps,superscriptaddress,showpacs,preprintnumbers]{revtex4} 
\usepackage{graphicx,epsfig,pstricks}
\usepackage{amssymb}
\usepackage{amsmath}
\usepackage{dsfont}
\usepackage{epstopdf}

\begin{document}

\title{The role of the chiral phase transition in modelling the kaon to pion ratio}

\author{A. V. Friesen}
\affiliation{Joint Institute for Nuclear Research,  Dubna,
Russia}

\author{Yu. L. Kalinovsky}
\affiliation{Joint Institute for Nuclear Research,  Dubna,
Russia}

\author{V. D. Toneev}
\affiliation{Joint Institute for Nuclear Research,  Dubna,
Russia}

\begin{abstract}
A sharp peak in the $K^+/\pi^+$ ratio in relativistic heavy-ion collision is discussed in the framework of the SU(3) Polyakov-loop extended NJL model with vector interaction.  In the model, the $K^+/\pi^+$ ratio was calculated along the chiral phase transition line for different values of the vector coupling $g_V$. We showed that the value of the vector coupling had no significant effect on the $K^+/\pi^+$ behaviour.
\end{abstract}


\maketitle

\section{Introduction}
Quantum chromodynamics predicts the existence of a quark-gluon plasma (QGP) phase at extremal conditions. In this phase, hadrons dissolve and quarks are supposed to be free and deconfined. The search for this phase in heavy ion collision is difficult due the short QGP lifetime. Various signals were proposed for detection of the QGP phase, and most of them are the matter of intense debate. It is supposed that the ''horn'', which appears in the ratio of positive charged kaon to pion, could be one of them 
\cite{NA49_exp,AndronicPLB}. 

Nowdays the picture of this peak becomes more clear from the experimental side, but still attracts many questions from the theoretical side. The peak appears in the ratio of positive charged kaons and pions at the collision energy $\sqrt{s_{NN}}\sim$ 7-10 GeV for the high-size systems in  Au+Au and Pb+Pb collisions \cite{NA49_exp,AndronicPLB}. With decreasing system size, the sharp peak becomes lower and for Be+Be,  p+p collisions the ratio demonstrates smooth behaviour \cite{NA61-SHINE}.

Many theoretical approaches can reproduce experimental data when the partonic phase is supposed to appear during the collision at low energies \cite{Bratkovskaya_chir,onset_deconf,CohenPRC,Nayak_PRC82}. The microscopic transport model with involving partial restoration of chiral symmetry at the early stages of the collision reproduces experimental data and predicts smoothing of the peak with decreasing system size \cite{Bratkovskaya_chir}. The authors showed that the partial chiral symmetry restoration was responsible for a quick increase in the $K^+/\pi^+$ ratio at low energies. It was explained that the decrease in the ratio after the maxima was a result of the chiral condensate destruction and QGP formation.

In this work, we discuss the chiral phase transition in the framework of the SU(3) NJL model with the Polyakov loop. In the model, the current quark propagating in the chiral condensate develops a quasi-particle mass that leads to spontaneous chiral symmetry breaking. 
At that time the quark is coupled to a homogeneous background field representing the Polyakov loop dynamics. Chiral symmetry is restored when the dynamically generated quark mass drops as a function of temperature and chemical potentials. At low chemical baryon potential and high temperature the chiral symmetry restoration is soft and is supposed to be a crossover. At high chemical potential and low temperature the chiral symmetry restoration is the first order phase transition. This picture can be changed in the PNJL model when the vector interaction is added. With increasing vector coupling, the domain of the first order transition decreases until it completely disappears. 

Varying the vector coupling, we can check if the change of the type of the phase transition affects the behaviour of the kaon to pion ratio in the low temperature and high chemical potential (low energy) region.

\section{ The PNJL model} 

In this work we used the SU(3) PNJL model with vector interaction. The NJL-like models are a practical tool for description of nuclear matter at finite T and $\mu_B$. The Lagrangian of the SU(3) PNJL model with vector interaction and $U_A(1)$ anomaly has the form \cite{U_Fu08, Role_Gv, SU3_vect}:  
\begin{eqnarray}
\mathcal{L\,} & = & \bar{q}\,(\,i\,{\gamma}^{\mu}\,D_{\mu}\,-\,\hat
{m} - \gamma_0\mu)\,q + \nonumber \\
&+& \frac{1}{2}\,g_{S}\,\,\sum_{a=0}^{8}\,[\,{(\,\bar{q}\,\lambda
^{a}\,q\,)}^{2}\,\,+\,\,{(\,\bar{q}\,i\,\gamma_{5}\,\lambda^{a}\,q\,)}%
^{2}\,] - \nonumber \\
&-& \frac{1}{2}g_{\rm V} \sum_{a=0}^{8}\,\,{(\,\bar{q}\gamma_\mu\lambda
^{a}\,q\,)}^{2} + \mathcal{L}_{\rm det} -  \mathcal{U}(\Phi, \bar{\Phi}; T) 
\label{lagr}%
\end{eqnarray}
where $q=(u,d,s)$ is the quark field with three flavours, and  $\hat{m}=\mbox{diag}(m_{u},m_{d},m_{s})$ is the current quark mass matrix, $\lambda^{a}$  are the Gell-Mann matrices, and D$_\mu = \partial^\mu -i A^\mu$, where  $A^\mu$   is the gauge field with $A^0= -  iA_4$ and $A^\mu(x) = g_SA^\mu_a\frac{\lambda_a}{2}$ absorbs the strong interaction coupling. The Kobayashi - Maskawa - t'Hooft (KMT) interaction is described by the term $L_{\rm det} = g_D \,\,\{\mbox{det}\,[\bar{q}\,(\,1\,+\,\gamma_{5}%
\,)\,q\,]+\mbox{det}\,[\bar{q}\,(\,1\,-\,\gamma_{5}\,)\,q\,]\,\}$. The confinement/deconfinement properties (Z$_3$-symmetry) are described by the effective potential $\mathcal{U}(\Phi, \bar{\Phi}; T)$, that is constructed on the basis of the Lattice inputs in the pure gauge sector (see for details \cite{EBlanquierJPG,Friesen2015}). In our work, we used the standard polynomial form of the effective potential \cite{Friesen2015,RattiPRD73}.

The grand potential density $\Omega(T,\mu_i)$ in the mean-field approximation can be obtained from the Lagrangian density (\ref{lagr}) and leads to the self-consistent equations:
\begin{equation}
\frac{\partial \Omega}{\partial \langle\bar{q_i}q_i\rangle}=0, \ \ \frac{\partial \Omega}{\partial \Phi}=0, \ \ \frac{\partial \Omega}{\partial \bar{\Phi}}=0, \ \ \frac{\partial \Omega}{\partial \rho_i}=0,
\end{equation}
from which we obtain the gap equation and the effective chemical potential: 
\begin{eqnarray}
m_i &=& m_{0i} - 2 g_S\langle\bar{q_i}q_i\rangle - 2g_D\langle\bar{q_j}q_j\rangle\langle\bar{q_k}q_k\rangle,  \\
\tilde{\mu_i} &=& \mu_{0i} - 2 g_{\rm V} \rho_i.
\end{eqnarray}
where $i, j, k = $u, d, s are chosen in cyclic order, $m_i$ are the constituent quark masses, the quark condensates and the quark density are given by:
\begin{eqnarray}
\langle\bar{q_i}q_i\rangle &=& - 2 N_c \int\frac{d^3p}{(2\pi)^3}\frac{m_i}{E_i}(1 - f^+_\Phi(E_i) - f^-_\Phi(E_i)) \\
\rho_i & =& 2N_c\int\frac{d^3p}{(2\pi)^3}(f^+_\Phi(E_i) - f^-_\Phi(E_i)).
\end{eqnarray}
with  the modified Fermi functions $f^\pm_\Phi(E_i)$: 
\begin{eqnarray}
\label{f-Phi}
f^+_\Phi(E_f)&=&
\frac{(\bar{\Phi}+2{\Phi}Y)Y+Y^3}{1+3(\bar{\Phi}+{\Phi}Y)Y+Y^3}
~,\\
f^-_\Phi(E_f)&=&
\frac{({\Phi}+2\bar{\Phi}\bar{Y})\bar{Y}+\bar{Y}^3}{1+3({\Phi}+\bar{\Phi}\bar{Y})\bar{Y}+\bar{Y}^3}
~,
\label{f-Phi-bar}
\end{eqnarray}
where the abbreviations $Y={\rm e}^{-(E_i-\mu_f)/T}$ and $\bar{Y}={\rm e}^{-(E_i+\tilde{\mu_i})/T}$ are used. The functions go over to the ordinary Fermi functions for $\Phi=\bar \Phi=1$.

At low temperature, chiral symmetry is spontaneously broken due to the
finite value of the quark condensate and confinement is observed in the system 
(the $Z_3$-symmetry is apparently restored). After reaching the critical temperature, the quark condensate drops to zero and partial chiral symmetry is restored. The Polyakov field $\Phi$ at critical temperature becomes nonzero, which means $Z_3$-symmetry breaking (or deconfinement). The chiral phase transition is defined as a local maximum of $\frac{\partial <\bar{q}q>}{\partial T}|_{\mu_B = const}$. The deconfinement transition is determined as maximum of $\frac{\partial \Phi}{\partial T}\vert_{\mu=const}$.

\begin{figure}[h]
\centerline{
\includegraphics[width = 0.5\columnwidth]{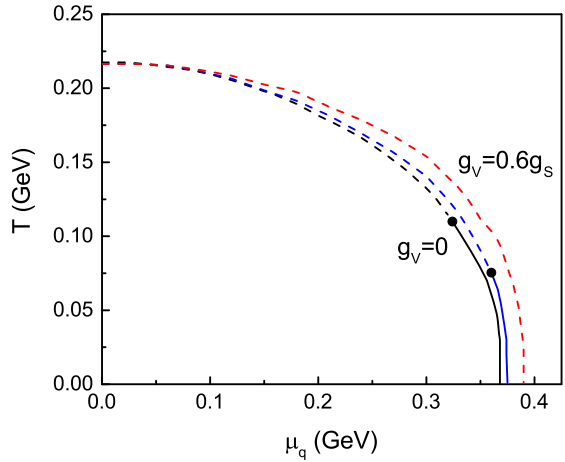}}
\caption{Phase diagrams in the T-$\mu_u$ plane for the cases with $g_{\rm V}=0$, $g_{\rm V}=0.2 g_S$ and $g_{\rm V}=0.6 g_S$. The dots show the critical end points for $g_V=0$ and $g_V=0.2g_S$.}
\label{3pd}
\end{figure} 

At low value of the chemical potential, the phase transition is smooth and the ''crossover'' means that the chiral condensate reduces in magnitude to half its value. At low temperatures and high chemical potential, the crossover turns into the first order transition. To describe the first order transition, the quark number susceptibility  $\chi_q=\frac{\partial^2 \Omega}{\partial \mu_u^2}|_{T = const}$ is introduced \cite{kunihiroPLB219}. The point where the crossover ends and the first order transition line starts is the second order phase transition critical end point (CEP).

In Fig.\ref{3pd}, the phase diagram in the T -$\mu_u$ plane for three choices of $g_{\rm V}$ is shown. In the case $g_{\rm V}$ = 0, the phase diagram has a familiar picture with the chiral crossover ending as a critical end point and the first order chiral phase transition extending from the CEP to the $\mu$ axis at T = 0. For $g_{\rm V}=0.2 g_S$, the phase diagram is similar, but the CEP moves down to to the $\mu$ axis. For $g_{\rm V} = 0.5g_S$ the first order phase transition has already disappeared and turned into a continuous crossover all the way down in temperature. In this work, we used the vector coupling $g_{\rm V} = 0.6 g_S$. It can be seen from Fig. \ref{3pd}, that increasing the vector coupling leads to the shift of the crossover pattern to higher chemical potentials.

We used a set of parameters: the current quark masses $m_{0 u} = m_{0 d} = 5.5$ MeV, $m_{0 s} = 0.131$ GeV,  the cut-off $\Lambda = 0.652$ GeV, couplings $g_D = 89.9$ GeV$^-2$ and $g_S = 4.3$ GeV$^{-5}$. We also introduced the strange chemical potential as $\mu_s = 0.5 \mu_u $ and $\mu_u = \mu_d$ \cite{FKT_PRC2019}. The model with the above set of parameters is characterized by the fact that the critical temperature of the crossover transition at $\mu_u = 0$ GeV is higher ($T_c = 0.216$) GeV than it was predicted by the Lattice QCD $T_c =154(9)$  \cite{Bazavov}.

\section{ Mesons in dense matter.}

Meson masses are defined by the Bethe-Salpeter equation at $\mathbf{P}=0$
\begin{equation}
1-P_{ij}\Pi_{ij}^{P}(P_{0}=M,\mathbf{P}=\mathbf{0})=0~, \label{rdisp}
\end{equation}
where for non-diagonal pseudo-scalar mesons $\pi\,,K$:
\begin{eqnarray}
P_{\pi}&=&g_{S}+g_{D}\left\langle\bar{q}_{s}q_{s}\right\rangle, \label{Ppi} \\
P_{K} &=& g_{S}+g_{D}\left\langle\bar{q}_{u}q_{u}\right\rangle. \label{Pkaon}
\end{eqnarray}
and the polarization operator has the form
\begin{equation}
\Pi_{ij}^{P}(P_{0})=4\left(  (I_{1}^{i}+I_{1}^{j})-[P_{0}^{2}-(m_{i}%
-m_{j})^{2}]\,\,I_{2}^{ij}(P_{0})\right). 
\label{ppij}
\end{equation}
Integrals  $I_{1}^{i}$ and $I_{2}^{ij}(P_{0})$ are defined as:
\begin{eqnarray}
I_{1}^{i}&=&iN_{c}\int\frac{d^{4}p}{(2\pi)^{4}}\,\frac{1}{p^{2}-m_{i}^{2}}, \nonumber\\
I_{2}^{ij}(P_{0})   &=& iN_{c}\int\frac{d^{4}p}{(2\pi)^{4}}\,\frac{1}%
{(p^{2}-m_{i}^{2})((p+P_{0})^{2}-m_{j}^{2})},\nonumber
\end{eqnarray}

\begin{figure}[h]
\centerline{
\includegraphics[width = 0.5\columnwidth]{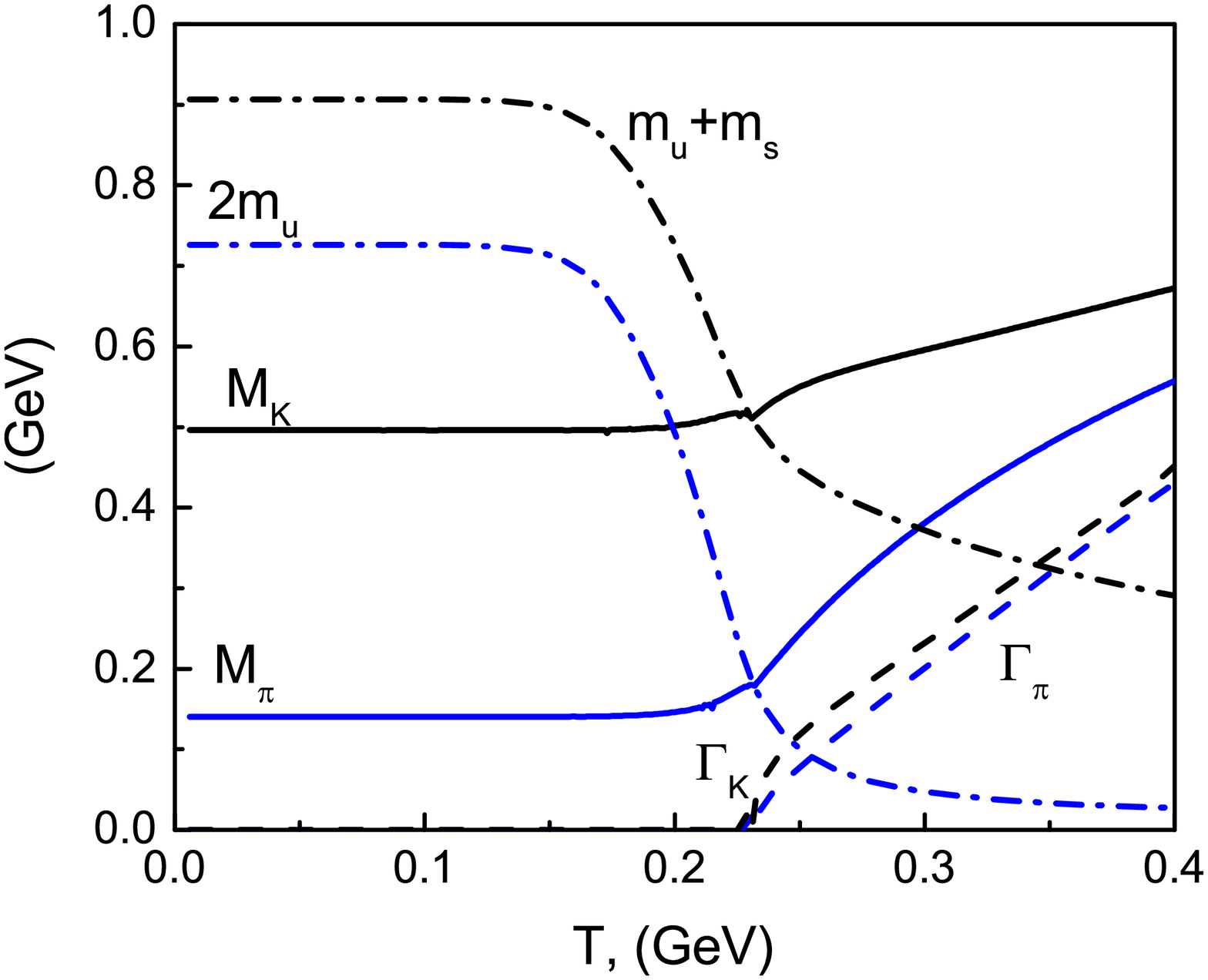}}
\caption{Meson masses as a function of temperature at $\mu_B = 0$.}
\label{masses}
\end{figure}

When the meson mass exceeds the sum of masses of its constituents ($P_0> m_i + m_j$), the meson turns into the resonance state and the Mott transition occurs. In this case, the complex properties of the integrals have to be taken into account and the solution has to be defined in the form ${P_0 = M_M - \frac{1}{2}i \Gamma_M}$. Equation (\ref{rdisp}) splits into two equations from which the meson mass $M_M$ and the meson width $\Gamma_M$ can be obtained. At zero chemical potential,  charged multiplets are degenerated and scalar mesons in the PNJL model decay at close temperatures: $T^\pi_{Mott} = 0.232$, $T^K_{Mott} = 0.23$ Gev (see Fig.\ref{masses}). 

\section{Results}
Though at the zero chemical potential pions and kaons are degenerated, their masses split with increasing density. At the nonzero chemical potential and low T, the mass splitting in charged multiplets due to excitation of the Dirac sea is modified by the presence of the medium \cite{FKT_PRC2019,Lutz,CostaKalin_2004}. In dense baryon matter the concentration of light quarks is very high \cite{Stachel_ss}. Therefore, the creation of a $s\bar{s}$ pair dominates because of the Pauli principle: when the Fermi energy for light quarks is higher than the $s\bar{s}$-mass, the creation of the last one is energy preferable.  The increase in the $K^+$  ($\bar{u}s$) mass with respect to that of $K^-$ ($\bar{s}u$) is justified again by the Pauli blocking for s-quark (see for discussion \cite{Lutz, Ruivo_1996}).
Technically, to describe mesons in dense matter, it is needed to  relate the chemical potential of quarks with the Fermi momentum $\lambda_i$,  $\mu_i = \sqrt{\lambda_i^2 + m_i^2}$. Pions are degenerated when masses of light quarks are equal ($m_u = m_d$) even at finite density.

All experimental data for the $K/\pi$ ratio were obtained in the midrapidity range. For the effective models, it means that the ratio of the particle number can be calculated in terms of the ratio of the number densities of mesons ($n_{K^\pm}/n_{\pi^{\pm}}$):
\begin{eqnarray}
n_{K^{\pm}} &=& \int_0^\infty p^2dp\frac{1}{e^{\beta(\sqrt{p^2+m_{K^{\pm}}}\mp\mu_{K^{\pm}})}-1}, \\
n_{\pi^\pm} &=& \int_0^\infty p^2dp\frac{1}{e^{\beta(\sqrt{p^2+m_{\pi^{\pm}}}\mp\mu_{\pi^{\pm}})}-1}.
\end{eqnarray}

The chemical potential for pions is a phenomenological parameter and in this work it was chosen as a constant $\mu_\pi = 0.135$ GeV, following the works \cite{pot_pi,BegunPRC90,naskret}. The chemical potential for kaons can be defined (see, for example, \cite{naskret,pot_K}) from $\mu_q = B_q\mu_B + S_q\mu_s + I_q\mu_q$, and in the  isospin symmetry case ($I_q = 0$), the result is $\mu_K =\mu_u-\mu_s$.

\begin{figure}[h]
\hspace{-0.7cm}
\centerline{
\includegraphics[width = 0.5\columnwidth]{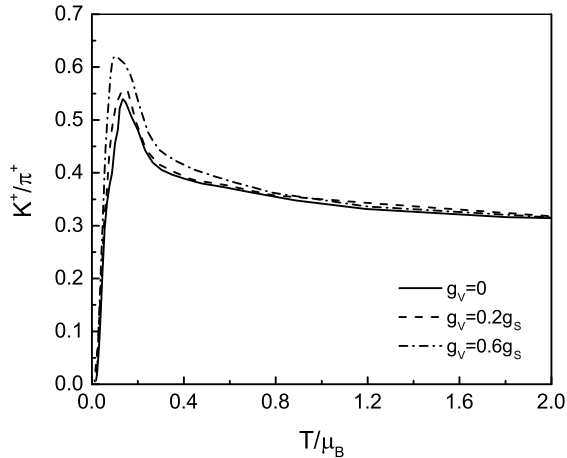}
}
\caption {The $K^+/\pi^+$ ratios as functions of the rescaled variable $T/\mu_B$ for different values of $g_{\rm V}$.  }
\label{nKnpi_ratio}
\end{figure}

There are some remarks to our calculations. The first is that the PNJL model is not able to describe the dynamics of the heavy ion  collision and it can be said that the model works at freeze-out. It is impossible to take into account mesons created at the latest stage of collision, out of equilibrium, appearing in the experiment naturally. The effective model calculation therefore pretends only to qualitative reproduction of data. The second point is that all experimental data are shown as a function of the collision energy $\sqrt{s_{NN}}$ which never appears as a parameter in effective models. To avoid this, we used the fact that in the statistical model for each experimental energy of collision the temperature and the baryon chemical potential of freeze-out can be found using parametrization suggested, for example, by J. Cleymans  \cite{Cleymans_diagr}. Using this parametrization, the $K/ \pi$ ratio can be considered as a function of a new variable $T/\mu_B$ instead of $s_{NN}$, where (T, $\mu_B$) are taken along the freeze-out line. In the mean field approach the chiral phase transition line can play the role of freeze-out, as in the PNJL model the chiral phase transition line divides the hadron phase and the quark-gluon phase. Therefore, in some approach, we can estimate the $K/\pi$ ratio along the phase diagram instead of using $\sqrt{s_{NN}}$. The results are shown in Fig. \ref{nKnpi_ratio} for different values of $g_{\rm V}$.

In this work, we  used the PNJL model with vector interaction to describe the QCD matter at finite temperature and density. It is well known that vector interaction leads to the situation when the first order transition in dense quark matter is replaced by the crossover and on the phase diagram there appears only a crossover line.  We paid attention to the kaon to pion ratio. All models, which were able to successfully describe the increase in the $K^+/\pi^+$ ratio, implied the phase transition (the chiral phase transition) at low energies. As $\sqrt{s_{NN}}$ can be replaced with the new variable $T/\mu_B$, where T and $\mu_B$ are calculated along the phase transition line, we can check if the increase of $g_V$ changes the $K^+/\pi^+$ behaviour.

We used the PNJL model with  $\mu_s = 0.5\mu_u$ and considered the cases with different values of $g_V$ which moves the CEP to lower T till it disappears and shifts the crossover pattern to higher chemical potentials. It can be seen in the Fig.\ref{nKnpi_ratio} that the absence of the first order phase transition domain leads only to a changing in the peak hight in the $K^+\pi^+$ ratio.

Using our previous study \cite{FKT_PRC2019,FKT_PEPAN19} we can conclude that the peak appears in the range of low temperatures and high baryon chemical potential (which corresponds to low energy of the collision).  The appearance of the peak is weakly sensitive to the type of phase transition in the high density region, as the replacement of the the first order transition to the soft crossover only leads  to a changing in the peak hight. The structure is more sensitive to the slope of the phase transition curve (see \cite{FKT_PEPAN19}) and the matter properties. For example, when the strange chemical potential is zero $\mu_S(\mu_K) = 0$, the $K^+/\pi^+$ ratio shows smooth behaviour. When the strangeness neutrality is introduced, the $K^+/\pi^+$ ratio does not show a peak structure \cite{FKT_PRC2019,FKT_PEPAN19}.

The work of A.F. and Yu. K. was supported by the RFBR, grant no. 18-02-40137.

\end{document}